\documentclass[onecolumn,           
               showpacs,            
               preprintnumbers,     
               aps,                 
               letterpaper,             
               superscriptaddress,      
               nofootinbib,         
               tightenlines,        
               floats,floatfix      
               ,usenatbib, 11pt, english
               ]{revtex4-1}
\renewcommand{\thesection}{\arabic{section}}
\renewcommand{\thesubsection}{\thesection.\arabic{subsection}}
\renewcommand{\thesubsubsection}{\thesubsection.\arabic{subsubsection}}

\usepackage{aas_macros}
\usepackage{graphicx}
\usepackage{subfigure}
\usepackage{hyperref}
\hypersetup{
    colorlinks=true,
    linkcolor=blue,
    filecolor=magenta,      
    urlcolor=cyan,
}
 
\usepackage{dcolumn}
\usepackage{bm}
\usepackage{listings}
\usepackage{floatrow}
\newfloatcommand{capbtabbox}{table}[][\FBwidth]

\usepackage{blindtext}
\usepackage{graphicx}  
\usepackage{dcolumn}   
\usepackage{bm}        
\usepackage{amsmath,amssymb}
\usepackage{hyperref}
\usepackage{color}
\definecolor{purple}{rgb}{0.58,0.0,0.83}
\usepackage{caption}
\usepackage[toc,page]{appendix}
\usepackage[utf8]{inputenc}
\usepackage{tabularx}
\usepackage[spanish]{babel}
\AtBeginDocument{\selectlanguage{english}}

\usepackage{booktabs}

\begin{document}
\title{Dark Matter with $N$-Body Numerical Simulations}
\author{Jazhiel Chac\'on}  
 \email{jchacon@icf.unam.mx}
\affiliation{Instituto de Ciencias F\'isicas, Universidad Nacional Aut\'onoma 
de M\'exico, Apdo. Postal 48-3, 62251 Cuernavaca, Morelos, M\'exico.}
\author{J.~Alberto V\'azquez}  
 \email{javazquez@icf.unam.mx}
\affiliation{Instituto de Ciencias F\'isicas, Universidad Nacional Aut\'onoma 
de M\'exico, Apdo. Postal 48-3, 62251 Cuernavaca, Morelos, M\'exico.}
\author{Ruslan Gabbasov} 
\email{ruslan.gabb@gmail.com}
\affiliation{Departamento de Sistemas, Universidad Aut\'onoma Metropolitana, Av. San Pablo 180, Col. Reynosa Tamaulipas, Alcaldía Azcapotzalco, C.P. 02200, CDMX, M\'exico.}

\date{\today}

\begin{abstract}

\subsection*{Abstract}
The development of numerical $N$-body simulations have allowed to study formation process and
evolution of galaxies at different scales. This paper presents the fundamental concepts 
of $N$-body systems applied to the cosmological evolution of the $\Lambda$-Cold Dark Matter ($\Lambda$CDM) model. 
In order to perform structure formation in the Universe, we provide an introduction to the 
basic equations and their implementation on the GADGET-2 software. We also present a 
simple guide to modify this code. 
First, we briefly describe the dark matter in the Universe as well as the theoretical and 
experimental basis of the $\Lambda$CDM model. Then, we focus on the simulation codes and 
provide the equations that govern most of the $N$-body simulations to model the dark matter. 
We describe the Smoothed Particle Hydrodynamics method used for simulating the gas, 
star dynamics and structure formation in these simulations. 
Then, cautiously, we guide the reader to the installation of GADGET-2 on a Linux-based computer, 
as well as to carry out a couple of examples to operate the code. 
Finally, by using a computational cluster, we show several results of a large structure
simulation, analyse the outputs to display the matter power spectrum, and compare the
outcome with theoretical predictions.

\textbf{\textit{Keywords}}---Numerical Simulations, $N$-body systems, cosmology, 
$\Lambda$CDM
\end{abstract}
 
\maketitle

\section{Introduction}

Over recent decades cosmology has played an important role in the development of science 
and technology. Its main goal seeks to explain the origin and evolution of the Universe 
as a whole, and hence, the fundamental physics behind those process to therefore gain a 
deeper understanding of the laws of physics \cite{2003itc..book.....R}. It is well known 
that the observable matter, galaxies, stars, gas clouds, planets and so on only contribute to 
about $\sim 5 \%$ of the total content of the Universe, whereas $27 \% $ corresponds to an 
unknown dark matter -- with the property to be gravitationally attractive -- responsible to be the main component 
of structure formation, and the remaining $\sim 68 \%$ corresponds to the dark energy -- the main 
candidate to explain the current  accelerated expansion of the Universe. These results conform 
the most well established model for the evolution of the Universe, the $\Lambda$-Cold Dark 
Matter model ($\Lambda$CDM).

Nonetheless, $\Lambda$CDM has been tested throughout the years using diverse experiments, just 
like the Planck mission and going back through other similar surveys, for example: the WMAP mission
results in 2007 \cite{2007ApJS..170..377S}, as well as the measured fluctuations in the CMB
temperature by the COBE satellite in 1994 \cite{1994ApJ...420..439M}. All of them, amongst 
many others, have contributed to reinforce the foundation of the $\Lambda$CDM model. 
One of the essential tests come from the $N$-body simulations, which are able to constrain 
several cosmological parameters. The key procedure of the $N$-body simulations is to evolve 
bound systems by considering dark matter interacts only gravitationally with ordinary
matter. This paper focuses on the basis for these kind of simulations in order to provide an
understanding of the cosmological evolution, through some commonly used codes.

The paper is structured as follows: First we provide a brief review about dark matter and 
its importance on the development of the $\Lambda$CDM model. Then, we present some numerical 
codes and their use on astrophysical systems, followed by the $N$-body simulations and their 
basic equations. We include a section dedicated to Smoothed Particle Hydrodynamics (SPH) and how 
to simulate gas dynamics. Later, we introduce the GADGET-2 software, a $N$-body-SPH hybrid 
free source code used in this work, and the basic installation procedure. Finally, by running the 
code, we present some results given different simulations, in particular from the evolution of 
two isolated galaxies colliding and merging into a larger galaxy, and also from the 
cosmological evolution of a periodic box described by a $\Lambda$CDM Universe. 
For the second example, we compute the matter power spectrum -an important quantity used to 
constrain physical quantities for a given cosmological model-. 
%

\subsubsection*{Dark Matter}

The beginning of dark matter history may be traced back to 1937 when 
the astrophysicist Fritz Zwicky examined the internal dynamics of the Coma Berenice galaxy cluster
\cite{1937ApJ....86..217Z}. In that work, Zwicky provided evidence that the luminous mass in the 
cluster was much smaller than the total mass needed to hold these galaxies together by gravitational
forces. Therefore, he concluded that there should be another type of matter that would allow galaxies 
to be gravitationally bounded. These observations were the first hints of a missing matter in 
galaxy clusters, -- the ``dark matter'' made its first appearance in the scientific community 
ever since --.

Despite numerous contributions from the scientific community, the issue of the dark matter was 
not seriously considered until the early years of the 1970 decade, when the astronomer Vera Cooper Rubin indicated that 
the gravitational stability of galaxies is due to an amount of mass greater than the observed
\cite{1970ApJ...159..379R}. In her work, she calculated the rotational curves of different 
spiral galaxies by measuring the radial velocity of the stars located at a distance
$r$ from the galaxy center, as seen in the following equation

\begin{equation} \label{eqn:1}
v (r)= \sqrt{\frac{G M (r)}{r}},
\end{equation} 
where $G$ is the Newton gravitational constant and $M(r)$ is the mass contained within the radius $r$.
According to Newton's laws, this movement is expected to be Keplerian, that is, the velocity of 
the stars would decline as the distance increases. The big surprise was that 
this curve does not follow the expected behaviour, as observations showed that the 
speed of the stars remained almost 
constant, and even in some cases it increased. If Newton's theory is correct, then a new kind 
of mysterious matter is needed whose mass distribution must increase with the radius. 
This strange behaviour is not observed on the Baryonic matter, which is distributed in a compact 
manner and its mass is not sufficient to maintain the flat rotation curve. 
The introduction of a new component caused a great impact on physics 
and astronomy, since it led to create alternative models that include dark matter in the galaxies, 
and therefore, also in the Universe.

Another evidence is the Cosmic Microwave Background (CMB) \cite{1965ApJ...142..419P} as being 
the earliest photograph of the Universe. The patterns seen on the CMB were set up by two competing 
forces acting upon matter: the gravity, causing the matter to fall inwards, and the radiation pressure,
preventing the gravitational collapse. This competition caused the photons and matter to oscillate 
in and out in dense regions forming patterns, that would be dramatically modified by the 
amount and type of dark matter present at that epoch.
That is,  the existence of dark matter leaves a characteristic imprint on CMB observations, as it 
clumps into dense regions and contributes to the gravitational collapse of matter, but it is 
unaffected by the pressure from photons.
The CMB power spectrum shows the strength of these oscillations at different scales, and for 
instance, the Wilkinson Microwave Anisotropy Probe (WMAP) \cite{2007ApJS..170..377S} was able to 
measure with enough accuracy the CMB  spectrum and consequently favoured the existence of dark matter.
 
Dark matter is also highly favoured when the Large Scale Structure formation is studied. 
The oscillations imprinted on the CMB evolved into more advanced
structures, given the amount of time available for objects to gravitationally collapse, eventually forming what is called the Baryon Acoustic Oscillations (BAO). 
At the time of CMB, dark matter did not undergo to the same oscillations with matter and light, 
but it was free to collapse on its own, this created dense regions that helped structure 
formation. Ths mechaninsm allowed the distribution of galaxies and clusters to be what it is observed today
\cite{2017AJ....154...28B}. 

\subsubsection*{Lambda Cold Dark Matter ($\Lambda$CDM)}

As mentioned above, one of the first predictions of the existence of dark matter was made by Zwicky. 
This result came from his observations on the Coma cluster in order to be able to explain 
its strange dynamics that would not match with a Newtonian behaviour. Although these observations 
were truly remarkable by that time, it wasn't until the 1980 decade that both astronomers and physicists
concluded that one way to explain the movement of the galaxies, according to Newtonian dynamics, 
was to include the ``missing'' matter predicted by Zwicky in the equations of motion.

The introduction of this missing matter (as well as the cosmological constant $\Lambda$) conforms 
the \textit{Lambda Cold Dark Matter} model. It is a parametrization to describe 
the cosmological Big Bang model and nowadays is referred as the ``standard cosmological model'', 
which is based on the following theoretical and experimental facts:

\begin{itemize}
\item A theoretical framework based on General Relativity, 
which provides a field theory for gravitation on cosmological scales.
\item The cosmological principle:  the Universe is
spatially isotropic and homogeneous on large scales \cite{2003itc..book.....R}.
\item The perfect fluid model:  the galaxies and the basic components of 
the Universe are included within the theory via the continuity equation \cite{schutz:2009}.
\item The Hubble's law establishes the expansion of the Universe in which the galaxies 
recession velocity is proportional to their distance \cite{1929PNAS...15..168H}.
\item The Cosmic Microwave Background radiation (CMB). The measurements of the CMB radiation 
support the cosmological principle on large scales 
\cite{1965ApJ...142..419P, 2007ApJS..170..377S, 2013IJMPD..2230029M}.
\item The determination of the relative abundance of primordial elements such as $^{1}$H, $^{2}$D,
$^{3}$He, $^{4}$He and $^{7}$Li, made up on nuclear reactions during the Big Bang 
Nucleosynthesis (BBN) era \cite{1993ppc..book.....P}.
\item The large scale structure analysis of the Universe using data from the Sloan Digital 
Sky Survey (SDSS) \cite{2017AJ....154...28B}, that aids the parameter determination of the 
standard cosmological model. 
\end{itemize}

Moreover, the $\Lambda$CDM model adds some other special features that allow to explain the 
evolution of the structure in the Universe:

\begin{itemize}
\item The evolution of matter density perturbations, initially coming from quantum density fluctuations, 
is required to explain the large scale structure in the Universe \cite{10.1143/PTP.76.1036}.
\item The Cosmic Inflation, originally introduced by Alan Guth, postulates an accelerated 
expansion at very early times that allows to explain the homogeneity and flatness in the Universe, 
as it is observed today \cite{1981PhRvD..23..347G}.
\item The cosmological constant $\Lambda$, introduced by Einstein on his equations of general 
relativity to force a static Universe. Nevertheless, it is known today that the Universe is 
in an accelerated expansion and this constant is referred to a form of vacuum energy or some 
kind of dark energy \cite{2003RvMP...75..559P}.
\item Cold Dark Matter (CDM). A sort of matter that has an exclusive gravitational attraction,  
does not interact with any kind of radiation (it is dark) and its velocity is not relativistic 
(it is cold).
\end{itemize}

\textbf{Some known issues with $\Lambda$CDM}\\

Although this model has been successfully proved by several observations and theoretical predictions,
 it has certain inconsistencies or unexplained features mainly on small scales.
Two of them are\\

\textbf{CUSP-CORE problem.} It refers to a discrepancy between the inferred dark matter density profiles of low-mass galaxies and the density profiles predicted by cosmological $N$-body simulations. Nearly all simulations with cold dark matter form halos which have ``cuspy"  distributions, with density increasing steeply at small radii, whereas the rotation curves of most observed dwarf galaxies suggest that they have flat central dark matter density profiles \cite{1996ApJ...462..563N, 1997ApJ...490..493N, Moore:1994yx}.

\textbf{Missing satellite problem.} It arises from a mismatch between observed dwarf galaxy numbers and numerical cosmological simulations that predict the evolution of the distribution of matter in the Universe. In simulations, dark matter clusters hierarchically, increasing the numbers of halo ``blobs" as halo components become smaller-and-smaller. However, there seem not to be enough observed normal-sized galaxies to match the simulated size distribution; the number of dwarf galaxies is orders of magnitude lower than expected from simulation\cite{1999MNRAS.310.1147M, 2005Natur.435..629S}.

With these deficiencies in mind, several alternative models have been suggested. 
Of a particular interest is to consider that the Dark Matter is made up of bosonic
excitations of an ultra-light scalar field minimally coupled to gravity, see \cite{Matos:2008ag, Magana:2012ph, Hui:2016ltb}
and references therein. We defer the numerical analysis with scalar fields  
for a future work.

\section{Numerical Codes for astrophysical systems}

To understand the large scale formation and structure of the Universe, the gravitational 
instability on cosmological scales and galaxy evolution, numerical $N$-body simulations are one 
of the most used approaches. Over recent years, the computational resources have allowed to create high
resolution simulations that recreate the evolution of the Universe since the CMB epoch ($z \sim 1100$).
Cosmological evolution is simulated with linear gravitational clustering on large scales 
($\geq 100 $ Mpc) and non-linear theory on small scales (between 10 kpc and 1 Mpc). On small scales,
specific initial conditions are created to evolve the dark matter particles, with the consideration 
that the dynamics can be enhanced by introducing effects of gas dynamics, chemical process, 
radiative transfer and other astrophysical phenomena.

There is a large variety of numerical codes that use the $N$-body theory and several 
applications including gas dynamics modelled by Smoothed Particle Hydrodynamics (SPH). These codes 
have been used in numerous times, and they have proved being a realistic approach according 
to observations. We list some of the methods used below:

\begin{enumerate}
\item \textbf{Direct methods}: these do not introduce approximations but they fully solve the
equations of motions and thus deliver the highest accuracy at the price of the longest computation
time, of order $\mathcal{O}(N^{2})$ per timestep. Integration is performed using adaptive
(individual) timesteps and commonly a fourth order Hermite integrator \cite{Mikkola1993}.
    
\item \textbf{Tree codes}: The tree code method (Barnes \& Hut 1986) provides a general 
integrator for collisionless systems. They take into account that particles nearby each other 
are important and the contributions from distant particles does not need to be computed 
with high accuracy, while potentials from distant groups of particles are approximated 
by multipole expansions about the group centres. The resulting computation time scales as
$\mathcal{O}(N\log(N))$ but the approximations introduce small force errors. The long-range 
force errors are controlled by a single parameter (the opening angle) that determines how small 
and distant a group of particles must be to use the approximation. Typical implementations of 
the tree code is to expand the potentials to quadrupole order and construct a tree 
hierarchy of particles 
using a recursive binary splitting algorithm. The tree does not need to be recomputed from 
scratch at every timestep, saving significant CPU time \cite{1986Natur.324..446B}.
    
\item \textbf{Particle-mesh codes}: This method is used as another approximation to speed up 
direct force calculation for collisionless systems. In this case the gravitational potential of 
the particular system is constructed over a grid starting from the density field and solving the
associated Poisson equation, by using Fast Fourier Transform. Particles do not interact directly 
amongst each other but only through a mean field. This method essentially softens the gravitational
interactions at small scales. The density field is constructed using a kernel to split the mass 
of the particles to the grid cells around the particle position. In a short range, accuracy of 
the force is a poor approximation of Newton's law up to several grid spacing distance 
\cite{klypin1997particlemesh}. 
    
\item \textbf{Adaptive Mesh Refinement method}: Particle-mesh codes can be enhanced by using 
an adaptive method rather than a static grid to solve the Poisson Equation. In the Adaptive 
Mesh Refinement (AMR) method the grid elements are concentrated where a higher resolution is needed, 
for example around the highest density regions. To obtain an adaptive resolution the method 
first uses a low-resolution solution of the Poisson equation, and then, progressively refining 
regions where a higher resolution is required \cite{2004astro.ph..3044O}. 
\end{enumerate}
%

\subsection{Basic $N$-body equations of motion}

It is well known that Einstein field equations describe the space-time behaviour in the 
presence of matter, that is 
\begin{equation}
    R_{\alpha \beta} - \frac{1}{2}R g_{\alpha \beta} + \Lambda g_{\alpha \beta} = 
        \frac{8 \pi G}{3}T_{\alpha \beta}, \label{eq: 2}
\end{equation}
where $R_{\alpha \beta}$ is the Ricci tensor, $R$ is the Ricci scalar, $g_{\alpha \beta}$ is the 
metric tensor, $\Lambda$ is the cosmological constant, and $T_{\alpha \beta}$ is the 
energy-momentum tensor. For a homogeneous and isotropic space-time, the energy-momentum tensor needs 
to be homogeneous and isotropic as well. This tensor is also known as a perfect fluid tensor and it 
has the following form
\begin{equation}
    T_{\alpha \beta} = diag(-\rho, p, p, p), \label{eq :3}
\end{equation}
inserting equation (\ref{eq :3}) into equation (\ref{eq: 2}) and setting $\Lambda = 0$, for a 
FRW metric with scale factor $a$, we obtain the Friedmann equations:
\begin{equation*}
    3 \frac{\dot{a}^{2} + k}{a^{2}} = 8\pi G \rho,
\end{equation*}
\begin{equation*}
    -2\frac{\Ddot{a}}{a} - \frac{\dot{a}^{2} + k}{a^{2}} = 8 \pi G p.
\end{equation*}
In the $\Lambda$CDM model, dark matter is assumed to be a non-baryonic matter component 
and its interaction is only gravitational, hence non-collisional. 
The $N$-body problem for these systems is described by the non-collisional Boltzmann equation in 
comoving coordinates coupled with the Poisson equation. A system of $N$ particles interacting
gravitationally defines a 6$N+1$ dimensional phase space given by the $N$ positions and 
velocity vectors associated to each particle at each time $t$. The solution of the $N$-body 
problem defines a trajectory in this phase space. On the other hand, if the number of particles 
is large enough, that is, if the two body relaxation time is long compared to the time-frame 
of interest, then a statistical description of the problem is possible. This allows to map 
the computation from a 6$N+1$ dimension to a 6+1 dimension phase space. The idea is to construct 
a mean field description of the dynamical system in terms of a single particle distribution function. 
The Boltzmann equation describes the behaviour and evolution of a fluid in the phase-space 
under external forces and has the following form 
\begin{equation}
\frac{\partial f}{\partial t} + \vec{v}\cdot\vec{\nabla}_{r}f + 
    \frac{\vec{F}}{m}\cdot\vec{\nabla}_{v} f=0,\label{eqn:2}
\end{equation}
where $f= f(\vec{r}, \vec{v}, t)$ is the distribution function of the density of the fluid, 
$\vec{v}$ is the velocity, $\vec{r}$ is the position, $\vec{F}$ is the force and $m$ is the mass 
of an individual particle of the system, that can describe eventually all the fluid.
If the force $\vec{F}$ is derived from a gravitational potential $\Phi$, it follows that
\begin{equation}
\vec{F} = -m\vec{\nabla}\Phi.\label{eqn:3}
\end{equation}
Substituting equation (\ref{eqn:3}) in (\ref{eqn:2}), it can be written as
\begin{equation}
\frac{\partial f}{\partial t} + \vec{v}\cdot\vec{\nabla}_{r}f -
    \vec{\nabla}\Phi\cdot\vec{\nabla}_{v}f=0.\label{eqn:4}
\end{equation} 
This potential $\Phi$ must satisfy Poisson's equation
\begin{equation}
\nabla^{2} \Phi (\vec{r},t) = 4\pi \int_{S} \int_{S}
    f(\vec{r},\vec{v},t)d^{3}\vec{v}d^{3}\vec{r},\label{eqn:5}
\end{equation}
where $S$ represents all space described by the total mass enclosed in a cube of volume 
$d^{3}\vec{r}$ centred in $\vec{r}$ and velocity $\vec{v}$ located in a cube of volume 
$d^{3}\vec{v}$ centred in $\vec{v}$. When integrating all over the space, the result is that 
the mass density may depend of time ($\rho(t)$), therefore, Poisson's equation described 
in equation (\ref{eqn:5}) can be reduced to a more familiar way.

Given its high dimensionality (6+1), the collisionless Boltzmann equation is usually solved 
by sampling the initial distribution function $f(\vec{r}, \vec{v}, t)$, and then, evolving 
the resulting $N$-body system, for instance with a numerical method that suppresses two body 
interactions at small scales. The interaction is softened not only for computational convenience 
to limit the maximum relative velocity during close encounters but especially to prevent 
artificial formation of binaries. 

In its discrete form, the Boltzmann equation describes the evolution of a set of point 
masses that auto interact gravitationally. In an $N$-body system, if $\vec{r}_{i}$ is the coordinate 
and $m_{i}$ is the mass of each particle, then Newton's equations of motion are
\begin{equation}
\frac{d^{2}\vec{r}_{i}}{d t^{2}}= -G \sum_{j=1, i \not= j}^{N} \frac{m_{j}(\vec{r}_{i}-
    \vec{r}_{j})}{|\vec{r}_{i}-\vec{r}_{j}|^{3}}, \label{eqn:6}
\end{equation}
using comoving coordinates $\vec{x}$ related with the physical coordinates $\vec{r}$ via the 
scale factor $a(t)$, it follows that $\vec{r} = a(t)\vec{x}.$
%
The evolution of the scale factor defines the Hubble factor
$    H(a) \equiv \frac{\dot{a}}{a}$,
through the Friedmann equation, as
\begin{equation}\label{eqn:8}
    H(a) = H_{0}[\Omega_{r, 0}a^{-4} + \Omega_{m, 0}a^{-3} + (1-\Omega_{0})a^{-2} +
        \Omega_{\Lambda, 0}]^{1/2}.
\end{equation}
$H_{0}$ $= 71 \pm 1$ km s$^{-1}$ Mpc$^{-1}$ is the Hubble's constant value at the present time,
$\Omega_{r, 0}, \; \Omega_{m, 0} \; \textup{and} \; \Omega_{\Lambda, 0}$ are the radiation, 
matter and dark energy densities, respectively, and their sum $\Omega = \Omega_{r} +
\Omega_{m} + \Omega_{\Lambda}$ must be one for a flat Universe.

In an expanding space modelled by a periodic box of size $L$,  Newton equations of motion 
can be deduced, in comoving coordinates, as
\begin{equation}
    \frac{d}{dt}(a^{2}\dot{\vec{x}}) = -\frac{1}{a}\nabla_{i}\phi(\vec{x_{i}}),\label{enq:10}
\end{equation}
\begin{equation}
    \nabla^{2}\phi(\vec{x}) = 4\pi G\sum_{i}m_{i}\left[-\frac{1}{L^{3}} +
    \sum_{\textup{\textbf{n}}}\delta(\vec{x} - \vec{x_{i}} - \textup{\textbf{n}}L)\right],\label{eqn:11}
\end{equation}
where the sum over $i$ is affecting the $N$ particles and $\phi$ is the peculiar gravitational potential
\begin{equation}
    \phi(\vec{x}) = \sum_{i}m_{i}\varphi(\vec{x}-\vec{x_{i}}),\label{eqn:12}
\end{equation} 
related to the Newtonian potential of a density fluctuation around a constant 
background density. The sum over the particles is also extended over their corresponding 
periodic images, with $\textup{\textbf{n}}=(n_{1},n_{2},n_{3})$ being a triple integral vector. 
The $-1/L^{3}$ factor is there to make sure that the mean density in Poisson's equation 
(\ref{eqn:12}) is different from zero, otherwise there would be no solution for an expanding 
space that tends to infinity. For a more detailed review, refer to the following reference
\cite{springel2014high}.

\subsection{Smoothed Particle Hydrodynamics (SPH)}

Smoothed Particle Hydrodynamics is needed to simulate astrophysical phenomena that involves
massive moving fluids in a 3-dimensional space. This method uses analytical differentiation 
with interpolation to compute the space derivatives, unlike the $N$-body approach, which divides 
the space into cells to compute the force between particles.  The SPH considers a set of 
discrete particles which represent the state of the fluid with continuous quantities associated 
to its motion, making the assumption that at any time, the position of the fluid elements are 
randomly distributed but the density is conserved. Obtaining the density is equivalent to obtaining 
the distribution probability of a fluid sample. An extended review of this topic can be found here
\cite{chacon_2019}.\\
The existing methods are:
\begin{itemize}
    \item Kernel softening. This method estimates the probability density function that describes the fluid \cite{parzen1962}.
    \item The spline delta technique. A differentiable curve defined by polynomials that allows the data analysis and aids the continuous modelling of the fluid \cite{1977MNRAS.181..375G, 1997JCoPh.136..298M}.
\end{itemize}

\subsection{Equations of motion}

The three fundamental equations are the energy density conservation equation, 
the momentum conservation equation and the Poisson equation. These can be in their 
integral formulation or in their differential form. The set of equations is called Navier-Stokes, 
and for fluids without viscosity they represent the Euler equations.
For cosmological simulations, the SPH approximation uses the perfect fluid model which is governed 
by the Euler equations of fluid dynamics, i.e. the continuity equation
\begin{equation}
 \frac{\partial\rho}{\partial t} + \vec{\nabla}\cdot(\rho\vec{v}) = 0,\label{eqn:13}
\end{equation}

 \noindent
and the momentum conservation equation
\begin{equation}
 \frac{\partial\vec{v}}{\partial t} + (\vec{v}\cdot\vec{\nabla})\vec{v} =
    -\frac{1}{\rho} \vec{\nabla}p - \vec{\nabla}\Phi, \label{eqn:14}
\end{equation}
along with the Poisson equation
\begin{equation}
 \nabla^{2}\Phi = 4 \pi G \rho,\label{eqn:15}
\end{equation}
where $\rho, \vec{v}, p$ are the density, velocity and pressure of the fluid at any time $t$. 
This set of equations gives a global view of the fluid. In Lagrange's representation, a point in 
the vector field is chosen at time $t = t_{0}$, and then, the temporal evolution is analysed, 
which allows to study the particle dynamics that make up the fluid individually. 
By expressing the total derivative as
\begin{equation}
 \frac{d}{dt} = \frac{\partial }{\partial t} +
 \vec{v}\cdot\vec{\nabla},\label{eqn:16}
\end{equation}
then, equation (\ref{eqn:13}) takes the following form
\begin{equation}
 \frac{d \rho}{d t} = - \rho \vec{\nabla}\cdot\vec{v},\label{eqn:17}
\end{equation}
and equation (\ref{eqn:14}) can be written as 
\begin{equation}
 \frac{d \vec{v}}{d t} =
 -\frac{1}{\rho}\vec{\nabla}p - \vec{\nabla}\Phi.\label{eqn:18}
\end{equation}
In order to describe a continuous fluid in a discrete approximation, 
SPH \cite{1992ARA&A..30..543M} starts by defining the integral interpolation of any 
function $A(\vec{r})$ as
\begin{equation}
 A_{I}(\vec{r}) = \int_{S} 
 A(\vec{r'})W(\vec{r}- \vec{r}', h)d^{3}\vec{r}',\label{eqn:19}
\end{equation} 
where the integration goes over all space and $W$ is an interpolation kernel that must satisfy
\begin{equation}
 \int_{S}W(\vec{r}- \vec{r}',h)d^{3}\vec{r}' = 1,\label{eqn:20}
\end{equation}
\begin{equation}
\lim_{h \to 0} W(\vec{r}-\vec{r}',h) = \delta(\vec{r}-\vec{r}').\label{eqn:21}
\end{equation}
The limit corresponds to the interpolation of the integral and $h$ is a length parameter 
in a 3-dimensional space. Numerical computations lead to a sum approximation
\begin{equation}
 A_{I}(\vec{r})=
 \sum_{j} m_{j} \frac{A_{j}}{\rho_{j}} W(\vec{r}- \vec{r_{j}},h),\label{eqn:22}
\end{equation}
where the index $j$ denotes each particle, and the sum is made over all the particles. 
Particle $j$ has mass $m_{j}$, position $\vec{r}_{j}$, density $\rho_{j}$, and velocity $\vec{v}_{j}$. 
Any other quantity $A$ inside $\vec{r}_{j}$ is denoted by $A_{j}$. The keypoint of this method 
is that it can build up a differentiable interpolator of any given function from its particular 
values (interpolation points) using an interpolation kernel that is also differentiable. 
There is no need of using finite differences or separating the space into cells just as $N$-body does. 
If it requires to compute $\vec{\nabla}A$, the calculation is simply
\begin{equation}
\vec{\nabla}A(\vec{r}) = 
    \sum_{j} m_{j} \frac{A_{j}}{\rho_{j}} \vec{\nabla}W(\vec{r}-\vec{r}_{j}, h).\label{eqn:23}
\end{equation}
The original calculations by Gingold \& Monaghan (1977) \cite{1977MNRAS.181..375G} uses an 
unidimensional gaussian kernel
\begin{equation}
  W(x,h)=
  \frac{1}{h \sqrt{\pi}} e^{-(x^{2}/h^{2})},\label{eqn:24}
\end{equation}
Nevertheless, to interpolate all over the nearest neighbours, a spline cubic function is usually defined as in the following reference (Springel et al. 2005 \cite{2005MNRAS.364.1105S})
\begin{equation}
    W(r,h) = \frac{8}{\pi h^{3}}\left\lbrace
\begin{array}{ll}
1 - 6(\frac{r}{h})^{2} + 6(\frac{r}{h})^{3},&  0\leq\frac{r}{h}\leq\frac{1}{2} \\\\
2 \left(1-\frac{r}{h}\right)^{3},&  \frac{1}{2} < \frac{r}{h}\leq1 \\\\
0,&\frac{r}{h}>1
\end{array}.
\right.\label{eqn:25}
\end{equation}
This is the usual example that mimics a delta function in the limit $h \rightarrow 0$. 
The choice of this kernel is such that the interaction recovers its Newtonian, original form 
at separations greater than the softening length  (See 2.4.1). For a physical interpretation 
of the SPH equations it is better to assume a gaussian kernel, for example, the density at any 
point in space is approximated by
\begin{equation}
\rho(\vec{r})=
\sum_{j} m_{j} W (\vec{r}-\vec{r}_{j},h).\label{eqn:26}
\end{equation}
Using this interpretation, the fluid density is now expressed in a discrete form by 
using the interpolation functions. By doing so, the continuity (\ref{eqn:13}), 
momentum conservation (\ref{eqn:14}) and Poisson (\ref{eqn:15}) equations pass from their continuous 
form to their discrete form, as described in reference  \cite{1985A&A...149..135M}.

\subsection{GADGET}

GAlaxies with Dark matter and Gas intEracT (GADGET), is a free source code which uses the 
$N$-body approach with SPH interpolation for cosmological simulations with its first version 
released in 2001 \cite{2005MNRAS.364.1105S}.
It is written in C language, and uses two main computational resources: Paralellization and 
the Tree--Particle Mesh Algorithm (TreePM). If a traditional method were used for computation purposes, 
it would require $N(N-1)$ force calculations for the $N$ particles, and the order of the 
computation time goes as $\mathcal{O}(N^{2})$. The TreePM method reduces the time to an order of 
$N \ln{N}$ by collecting all the particles within a cube of a given minimum size, together 
with paralellization to allow the system of millions of particles be computed in a more efficient 
way without losing much resolution.
%

\subsubsection{Gravitational softening} \label{GravSoft}
Because the large number of particles and information managed by the $N$-body simulations, 
if two particles are really close in space, that would lead to a divergence in the force acting 
upon a pair of particles. In order to avoid this divergence and exceeding accelerations, if 
two particles are very close to each other, a gravitational softening is introduced which must 
be acting on the whole space of the simulation. 
This gravitational softening is there to prevent that particles within the simulated box 
come very close to each other, in other words, the gravitational softening acts as a constriction 
for the simulated particles, also allowing the particles to remain in the non-collisional regime 
needed to solve the Boltzmann equations. 
This is achieved by introducing a parameter $\epsilon^{2}$ into equation (\ref{eqn:6}), as follows
\begin{equation}
\frac{d^{2}\vec{r}_{i}}{d t^{2}}=
    -G \sum_{j=1, i \not= j}^{N} \frac{m_{j}(\vec{r}_{i}-\vec{r}_{j})}{(\Delta\vec{r}_{ij}^{2} +
        \epsilon^{2})^{3/2}},\label{eqn:27}
\end{equation} 
where $\Delta\vec{r}_{ij}^{2} = |\vec{r}_{i} - \vec{r}_{j}|^{2}$ and $\epsilon$ is the softening 
length (Bodenheimer et al., 2007 \cite{2007nmai.conf.....B}). The physical interpretation of 
$\epsilon$ is the distance between the two centers of two ``binded'' particles. There is no criteria 
for the choice of the value of $\epsilon$, but for non-collisional systems, numerical results 
suggest using the mean separation between particles as a reference, although, it really depends 
on the size of the system that is being computed.
%

\subsubsection{Tree--Particle Mesh algorithm and Paralellization}

The tree particle method (Barnes \& Hut 1986) \cite{1986Natur.324..446B} provides a fast, 
general integrator for collisionless systems, when close encounters are not important and where 
the force contributions from very distant particles do not necessarily need to be computed with 
high accuracy. In fact, with a tree code, at small scales, strong interactions are typically 
softened, while the potentials due to distant groups of particles are approximated by 
multipole expansions about the group centres of mass. The resulting computation time scales as
$\mathcal{O}(Nlog(N))$, but the approximations introduce small force errors. The long-range force 
errors are controlled by a single parameter (the opening angle) that determines how small and 
distant a group of particles must be to use the approximation. This strategy works well to keep 
the average force error low.

\begin{figure}[t]
    \centering
      \subfigure[BH]{\includegraphics[width=0.45\textwidth, height=0.4\textwidth]{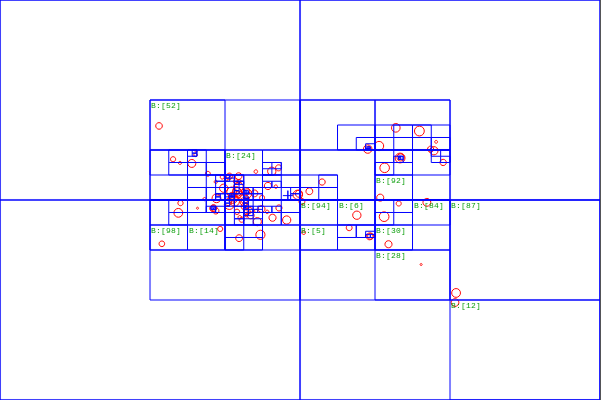}}
      \subfigure[TreePM]{\includegraphics[width = 0.4\textwidth]{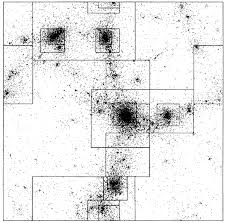}}
      \caption{\footnotesize{\textit{Left} Barnes \& Hut (BH) algorithm for 100 particles.
      \textit{Right}: The density field is interpolated over the nested mesh to ease the force
      interaction calculation between particles and their gravitational potential.}}
    \label{fig1}
\end{figure}

On the other hand, the concept of the Tree-PM algorithm is that a large number of particles or 
bodies can be approximated by a very well defined mesh that has the properties of the particles 
as a whole. They are organised in a branched system where the ``root'' has the complete information 
of the $N$-body system. The density field of the simulation is divided into cubic cells, in which if any 
cell has no information (has no particles), this cell is put aside, and if the cell has at least 
one particle, the force calculations begin and the cell now becomes a node. 
Each node is divided into $8$ cubes recursively until only one particle is left and the algorithm 
stops (Figure (\ref{fig1})).\\

\textit{Paralellization}: In order to perform calculations, GADGET distributes the volume 
of the simulation all over the processors of the computer in a Peano-Hilbert 
curve \cite{2005MNRAS.364.1105S}. This curve carries all particle information and divides it 
gradually in the processors which allows an equally distributed load.
\\
\textit{Force interaction computation}: For each particle, the Tree-PM algorithm produces a branch 
from the computer root. In this case, the root is the main node and behaves as a mesh that 
spreads on the next nodes. If the current node is at a smaller distance from the particle at 
which the calculation is being made, then the node is added to an interaction list.

If, on the other hand, the centre of mass of the node is at a greater distance from the particle, then the node opens and the following question is made: Is the distance of the centre of mass 
of the mesh greater than the size of the initial cube divided by some parameter, say $0 < \theta \leq 1$?
In other words, the question is whether the following relation is fulfilled
\begin{equation}
    r > \frac{l}{\theta}, \label{eqn:28}
\end{equation}
where $r$ is the distance of the particle to the centre of mass of the mesh, $l$ is the size 
of the initial cube and $\theta$ is a precision parameter. If the expression (\ref{eqn:28}) is 
true for every particle in the simulation, it continues evolving; on the other hand, if one or more particles 
do not satisfy this relation, the initial cube is divided into a smaller cube of size $l/2$ 
and the process is repeated. The algorithm computes multipole expansions and allocate the centre 
of mass of each cube. After that the question is asked again for each process (Figure \ref{fig1}).

Alternatively, the SPH method for cosmological simulations is mainly used for modelling the 
interstellar medium and uses the information obtained previously from the force interaction, 
like the nearest neighbours list to the particle and the force between them. By doing so, 
the computing time is drastically reduced by avoiding loop calculations.

\subsection{GADGET-2 Installation}
GADGET-2 is a free source code, ready to download and use it for making simulations. 
The instructions to compile and run the code may depend on the operating system that is being used; 
if the user has a Lunix or UNIX based system, the compilation is very straightforward. 
In the case of a MacOS user,
an \textsf{Xcode} update is needed, which includes all necessary compilation tools to install the code.
For a Windows system, the environment \textsf{Cygwin} is required to use any UNIX based 
compilation system.
\\

The following software is required:

\begin{enumerate}
\setlength
\itemsep{0em}
\item \hyperlink{https://wwwmpa.mpa-garching.mpg.de/gadget/gadget-2.0.7.tar.gz}{\underline{Gadget-2.0.7}}.
\item A \underline{1.9} or higher version of \hyperlink{https://www.gnu.org/software/gsl/}{\underline{GNU scientific library}} (GSL).
\item The \underline{2.1.5} version of  \hyperlink{http://www.fftw.org/fftw-2.1.5.tar.gz}{\underline{FFTW fast Fourier Transform in the West}}.
\item A parallel processing library, like Message Passing Interface (MPI) or \hyperlink{https://www.open-mpi.org}{\underline{OpenMPI}} or \hyperlink{https://www.mpich.org}{\underline{MPICH}}.
\item THe HDF format library dependencies \hyperlink{https://support.hdfgroup.org/ftp/HDF5/prev-releases}{\underline{Hierarchical Data Format}}, versión 1.6.10.
\end{enumerate}
The parallel processing libraries can be directly installed on a Linux based system. 
The \textsf{OpenMPI} package comes within the MacOS systems. On a side note, DO NOT download 
any \underline{3.x} version of FFTW, because it does not support parallel processing.

Once the software is downloaded, proceed to unzip the \textsf{.tar.gz} file and install. 
The following process is made on a Linux terminal, so be aware of that:

\begin{enumerate}
\setlength
\itemsep{0em}
\item Extract the software:

\textsf{user@PC$\sim$/Documents/code: tar -xzvf fftw-2.1.5.tar.gz}

\textsf{user@PC$\sim$/Documents/code: tar -xzvf gsl-1.9.tar.gz}

\textsf{user@PC$\sim$/Documents/code: tar -xzvf gadget-2.0.7.tar.gz}

\textsf{user@PC$\sim$/Documents/code: tar -xzvf hdf5-1.6.10.tar.gz}

\item Install GSL:
\textsf{user@PC$\sim$/Documents/code: cd gsl-1.9/}

\textsf{user@PC$\sim$/Documents/code/gsl-1.9: ./configure}

\textsf{user@PC$\sim$/Documents/code/gsl-1.9: make}

\textsf{user@PC$\sim$/Documents/code/gsl-1.9: sudo make install}

This is a \textsf{root} installation. It may depend on the computer manager to give admin privileges 
to the user or in other case, to install it on another folder, making sure the path to 
the necessary libraries is correct.
\begin{verbatim} --prefix=/path/to/folder/ \end{verbatim}
\item Install FFTW:

\textsf{user@PC$\sim$/Documents/code: cd fftw-2.1.5}

\textsf{user@PC$\sim$/Documents/code/fftw-2.1.5: ./configure - -enable-mpi - -enable-type-prefix - -enable-float}

\textsf{user@PC$\sim$/Documents/code/fftw-2.1.5: make}

This step takes roughly 10 minutes, so feel free to go for a coffee or a snack. Finally, install the libraries as root:

\textsf{user@PC$\sim$/Documents/code/fftw-2.1.5: sudo make install}

\item Install the HDF library:

\textsf{user@PC$\sim$/Documents/code: cd hdf5-1.6.10}

\textsf{user@PC$\sim$/Documents/code/hdf5-1.6.10: ./configure}

\textsf{user@PC$\sim$/Documents/code/hdf5-1.6.10: sudo make install}

\item Edit the Gadget \textsf{Makefile}:

This code has a wide variety of parameters to compile which are richly described on the User's guide.
Inside the Gadget's compile folder go to the \textsf{Gadget-2} folder and open the \textsf{Makefile} 
in a terminal or a notepad, then edit the \textsf{Makefile} to follow the path where the 
GSL and FFTW libraries were installed:
\footnotesize{
\begin{verbatim}
#-------Adjust settings for target computer
.................
#HDF5INCL =  
#HDF5LIB  =  -lhdf5 -lz 
endif
\end{verbatim}
}
\end{enumerate}
By default, the libraries are located in \textsf{/usr/local/}. The installation of Gadget 
in a computational cluster is a little bit tricky, please take a look to this document made by 
HPC Advisory Council{\footnote{\url{www.hpcadvisorycouncil.com/pdf/GADGET-2_Best_Practices.pdf}}}. 
The \textsf{Makefile} has to be edited depending on the system that will be simulated.

\section{Examples}
Before executing any simulation, the \textsf{Makefile} inside the \textsf{Gadget-2} folder needs 
to be edited. These two following examples are two different systems: a) two colliding disk galaxies 
and b) the large scale structure formation in a $\Lambda$CDM Universe; and they were run in a  
4-cpu computer.

\subsection{Colliding galaxies}

\begin{figure}[t]
    \centering
    \includegraphics[scale = 0.27]{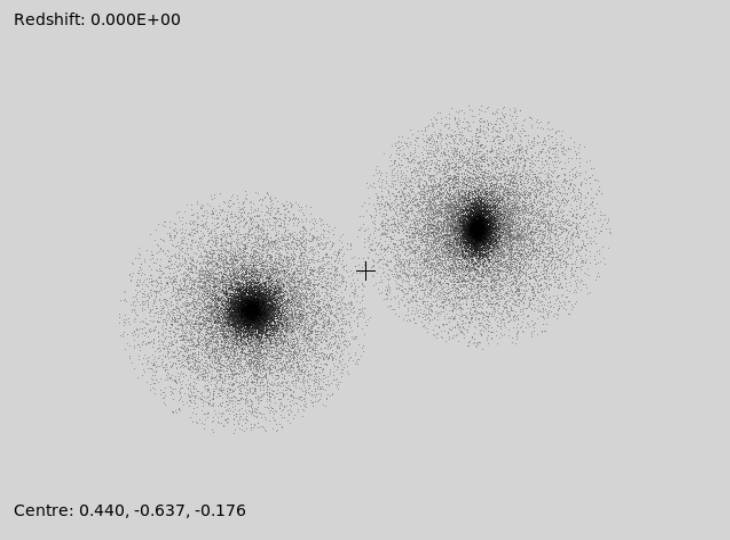}
    \centering
    \includegraphics[scale = 0.3]{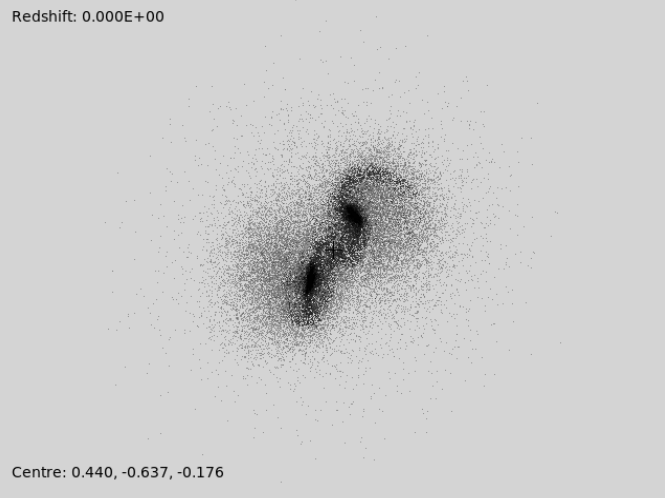}
    \centering
    \includegraphics[scale = 0.27]{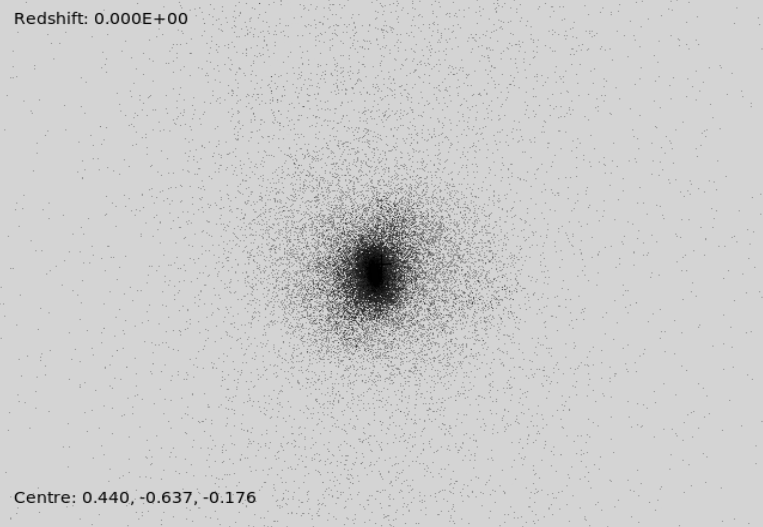}
    \caption{\footnotesize{Two disk galaxies colliding. The disk has 20000 particles 
    and the dark matter halo has 40000 particles. The galaxies are initially placed close to 
    each other with an axisymmetric disk and are attracted by their gravitational force. 
    They collide forming a pair of spiral perturbed galaxies to finally merge in just one 
    elliptical galaxy.}}
    \label{fig4}
\end{figure}

This simulation consists of two disk galaxies approaching each other, leading to a fusion 
between them. Each galaxy has a stellar disk and a dark matter halo using Newtonian Physics, 
with 20,000 disk particles and 40,000 dark matter halos (Figure \ref{fig4}). For this example, 
the following lines of the \textsf{Makefile} are modified:
\footnotesize{
\begin{verbatim}
    
#--------------------------------------- Basic operation mode of code
#OPT   +=  -DPERIODIC 
OPT   +=  -DUNEQUALSOFTENINGS

#--------------------------------------- Things that are always recommended
OPT   +=  -DPEANOHILBERT
OPT   +=  -DWALLCLOCK   

#--------------------------------------- TreePM Options
#OPT   +=  -DPMGRID=256
..................
\end{verbatim}
}

\normalsize{}
The rest of the file remains the same.  
\noindent
To run this simulation, it is highly recommended to make a new working folder with the parameter 
files and executables in order to avoid eventual troubles because of reediting 
the \textsf{Makefile}

\noindent
\textsf{user@PC$\sim$/Documents/code/Gadget-2.0.7: mkdir galaxy}\\

\noindent
Then, copy the \textsf{.exe} file in the \textsf{galaxy} folder:\\

\noindent
\textsf{user@PC$\sim$/Documents/code/Gadget-2.0.7: cp Gadget2/Gadget2 galaxy/}\\

\noindent
The parameter files have all the information that the simulation needs to run: the particle number, 
the initial conditions and so on. This file is inside the \textsf{parameterfiles folder}, 
make sure to copy them to the \textsf{galaxy} folder:\\

\noindent
\textsf{user@PC$\sim$/Documents/code/Gadget-2.0.7: cp Gadget2/parameterfiles/
galaxy.param galaxy/}\\

\noindent
Now, edit the default parameter file named \textsf{galaxy.param}\\

\noindent
\textsf{user@PC$\sim$/Documents/code/Gadget-2.0.7: cd galaxy}\\
\noindent
\\
The two first lines have to look as follow:
\footnotesize{\begin{verbatim}
% Relevant files

InitCondFile /path/to/Gadget-2.0.7/ICs/galaxy_littleendian.dat
OutputDir /path/to/Gadget-2.0.7/galaxy/
\end{verbatim}
}
\normalsize{
where \textsf{path/to/Gadget-2/ICs} and \textsf{path/to/Gadget-2/galaxy} are the paths 
where the Initial conditions are being read and where the output files want to be created. 
Now all is set to run the first collision simulation:\\}

\noindent
\textsf{user@PC$\sim$/Documents/code/Gadget-2.0.7/galaxy: mpirun -np 2 ./Gadget2 galaxy.param}\\

\noindent
This line calls for an MPI script to parallel processing. The \textsf{-np 2} indicates how 
many processors will be used for the computation.

\subsection{Large scale structure formation}
\begin{figure}[t]
    \centering
    \includegraphics[trim = 1mm  1mm 1mm 1mm, clip, width=5.cm, height=4.5cm]{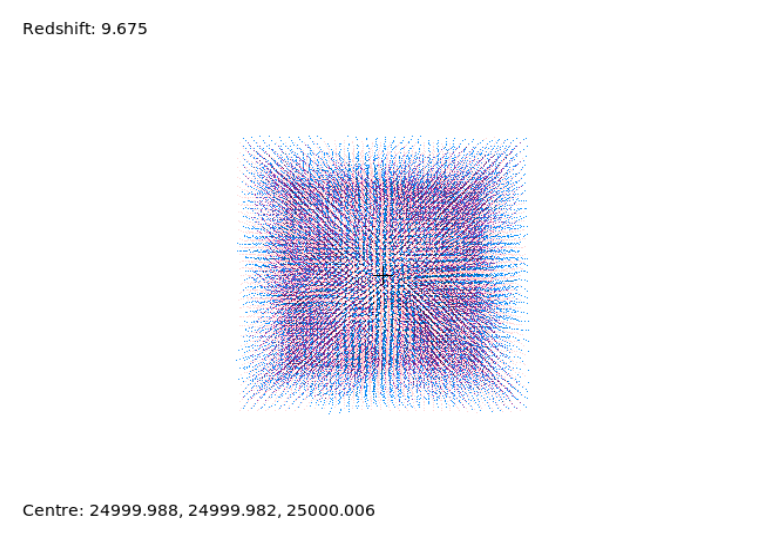}
        \includegraphics[trim = 1mm  1mm 1mm 1mm, clip, width=5.cm, height=4.5cm]{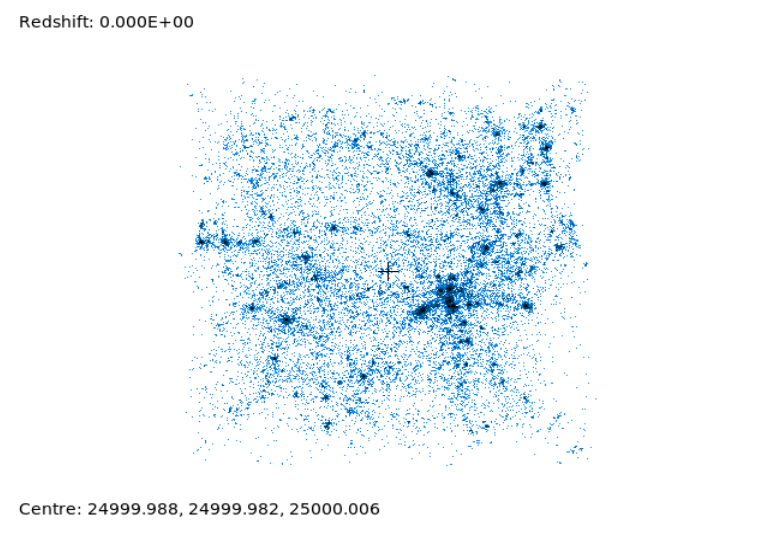}
    \includegraphics[trim = 1mm  1mm 1mm 1mm, clip, width=5.cm, height=4.5cm]{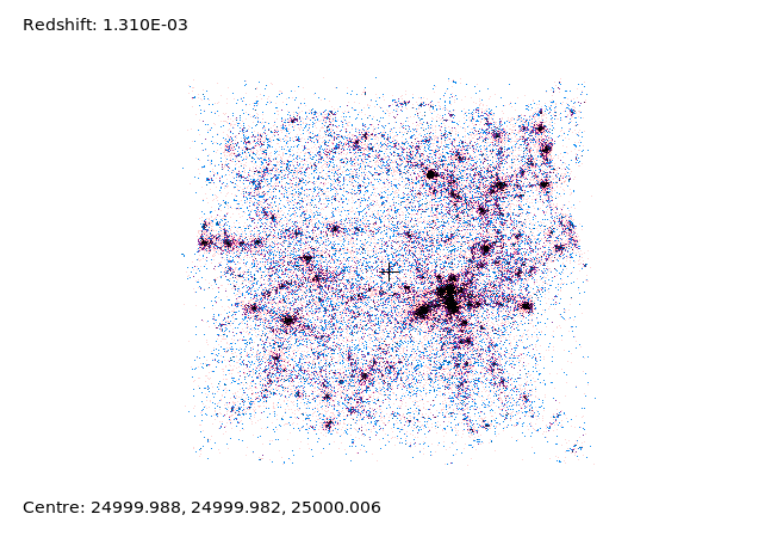}
    \caption{\footnotesize{Large scale structure formation. The simulation begins 
    by placing the particles in a cubic mesh, a perturbation makes the particles to evolve and 
    then they form galaxy clusters. Blue particles represent dark matter and red particles gas.}}
    \label{fig5}
\end{figure}

This is an example of $32^{3}$ dark matter particles and $32^{3}$ 
gas particles. The structure formation is made within a periodic box of size $50 h^{-1}
\textup{Mpc}$ per side in a $\Lambda$CDM Universe (Figure \ref{fig5}). This simulation 
distributes the particles in a cubic mesh, where they are placed in the mesh centres surrounded 
by dark matter particles. A perturbation on the position makes the particles move and eventually 
they form structures. The code starts running from $z = 10$ and finishes at the present epoch
($z = 0$). 
The parameters of this simulation are indicated in Table \ref{Tabla 3.1}.

\begin{table}[htb]%
\caption{Parameters of a $\Lambda$CDM simulation with Gadget}
\label{Tabla 3.1}%
\centering
\begin{tabularx}{0.9\textwidth}{@{\extracolsep{\fill}}  l c c }
\toprule%
Description & Symbol & Value\\\toprule%
Dark matter density & $\Omega_{0}$ & 0.3\\
Dark energy density & $\Omega_{\Lambda}$ & 0.7\\
Baryonic matter density & $\Omega_{b}$ &0.04\\
Hubble parameter & $h$ & 0.7\\
($h=H_{0}/100$ $\textup{Mpc}\cdot\textup{km}\cdot\textup{s}^{-1}$)\\ 
\hline
\end{tabularx}

\end{table}

The Gadget \textsf{Makefile} needs to be edited as follows:

\footnotesize{
\begin{verbatim}
#--------------------------- Basic operation mode of code
OPT   +=  -DPERIODIC 
#OPT   +=  -DUNEQUALSOFTENINGS

#------------------------------- Things that are always recommended
OPT   +=  -DPEANOHILBERT
OPT   +=  -DWALLCLOCK   

#----------------------------------------- TreePM Options
OPT   +=  -DPMGRID=128
......................
\end{verbatim}
}
\normalsize{}
To run the software it is necessary to call the parameters file \textsf{lcdm\_gas.param}. 
A variety of codes to generate initial conditions for large scale structure formation exists, 
these codes use Lagrangian Perturbation theory (LPT) such as the Zeldovich Approximation (ZA). 
For this example, the initial conditions were created using the \textsf{N-GenIC} software which 
can be easily manipulated. To run the code, just execute it as follows:\\

\noindent
\textsf{user@PC$\sim$/Documents/code/Gadget-2.0.7/lcdm\_gas:} \textsf{mpirun -np 2/   ./Gadget2 lcdm\_gas.param}\\

This example takes roughly 20 minutes to finish. This is because the number of particles 
simulated are just a few compared to a major resolution simulation, in which case it needs to 
be executed on a computational cluster.

\section{Creating initial conditions}

GADGET is a code that evolves a system of particles, the initial conditions need to be 
created using other codes and resources such as GalIC \cite{2014MNRAS.444...62Y} and N-GenIC 
\cite{2009MNRAS.394.1559G}, to create initial conditions for galaxies and large 
structures respectively. Other codes can also be used for such purposes, as MusIC
\cite{2011MNRAS.415.2101H}
and 2LPTIC \cite{2006MNRAS.373..369C} that use a Second Order Lagrangian Perturbation Theory.

\subsection{GalIC}
This code uses an iterative method to compute $N$-body simulations in equilibrium systems 
given its density distributions, such as spherically symmetric functions, axisymmetrical systems 
and galaxy models with different density profiles. There are two versions of GalIC, the 
\textsf{galic 1.0} version and the \textsf{galic 1.1} version. The installation of the first version 
is quite similar to installing GADGET. For the 1.1 version, it is also necessary a Doxygen tool, 
because this version is intended to be more accessible to another operating systems and 
programming languages.

The folder includes a list of examples listed on Table 1 of reference \cite{2014MNRAS.444...62Y}, and the parameters may be changed to make major resolution simulations.

\subsection{N-GenIC}
This code uses the Zeld\'ovich Approximation \cite{1970A&A.....5...84Z, RevModPhys.61.185}, 
which describes a non-linear evolution of the state of a matter density gravitational perturbation, 
which is considered to be homogeneous and non-collisional. In the file, the following parameters 
can be edited:

\begin{enumerate}
\setlength
\itemsep{0em}
\item Simulation including either only dark matter or dark matter with gas particles.
\item The number of particles $N$.
\item The initial time of the simulation $z_{i}$.
\item Dark matter density ($\Omega_{0}$).
\item Dark energy density ($\Omega_{\Lambda}$).
\item Baryonic matter density  ($\Omega_{b}$).
\item Hubble's parameter ($h$).
\item Boxsize of the simulation ($L$).
\item Power spectrum normalization ($\sigma_{8}$) \cite{1993MNRAS.262.1023W}.
\end{enumerate}

\section{Results}

Using the parameters of Table \ref{Tabla 4.1}, a structure formation simulation was 
carried out starting from $z = 23$ to $z = 0$ in a $\Lambda$CDM Universe. On the other hand,  
the matter power 
spectrum was generated with the code CAMB \cite{2000ApJ...538..473L}, and compared with the 
outcome from the simulation (see Figure \ref{fig7}).

\begin{table}[t]%
\caption{Initial conditions}
\label{Tabla 4.1}\centering%
\begin{tabularx}{0.9\textwidth}{@{\extracolsep{\fill}}  l l c c }
\toprule%
&Description&Symbol&Value\\\toprule%
Densities at $z=z_{f}$& Dark matter&$\Omega_{0}$&0.268\\
&Dark energy&$\Omega_{\Lambda}$&0.683\\
&Baryonic matter&$\Omega_{b}$&0.049\\\midrule
Simulation&Boxsize&$L$& 50 Mpc\\\
&No. of particles&$N$& 4096$\times 
12^{2}$\\\midrule
Redshift&Initial&$z_{init}$& 23\\
&Final&$z_{f}$&0\\\midrule
Other quantities&Hubble's parameter&$h$&0.7\\
&Matter power spectrum normalisation&$\sigma_{8}$&0.8\\\bottomrule
\end{tabularx}
\end{table}

\begin{table}[t]%
\caption{Additional parameters}
\label{Tabla 4.2}\centering%
\begin{tabularx}{0.9\textwidth}{@{\extracolsep{\fill}}  l l c c }
\toprule%
&Description&Quantity&Units\\\toprule%
Unit system&Length(cm)&$3.08\times10^{21}$&1 kpc\\
&Masa (g)&$1.989\times10^{43}$&$10^{10}$ $M_{\odot}$\\
&Velocity (cm/s)&$10^{5}$&1 km/s\\\midrule
Softening&$\Lambda$CDM($\epsilon$)&0.89, 20 &kpc\\\bottomrule

\end{tabularx}
\end{table}

\begin{figure}
    \centering
        \includegraphics[trim = 1mm  1mm 1mm 1mm, clip, width=9.cm, height=7.cm]{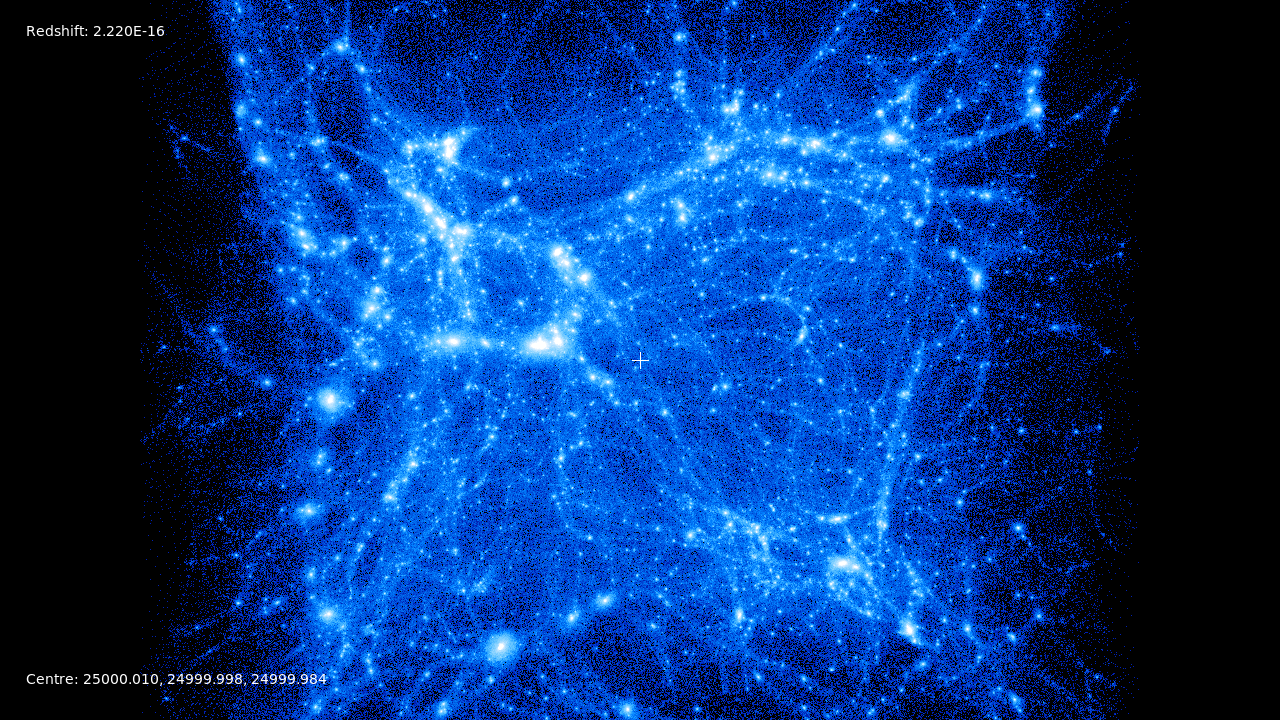}
    \caption{\footnotesize{A front slice of a 3D view of the final output of the simulation. Only dark matter particles were evolved.}}
    \label{fig6}
\end{figure}

\begin{figure}[h!]
    \centering
        \includegraphics[trim = 1mm  0mm 1mm 1mm, clip, width=12.cm, height=8.cm]{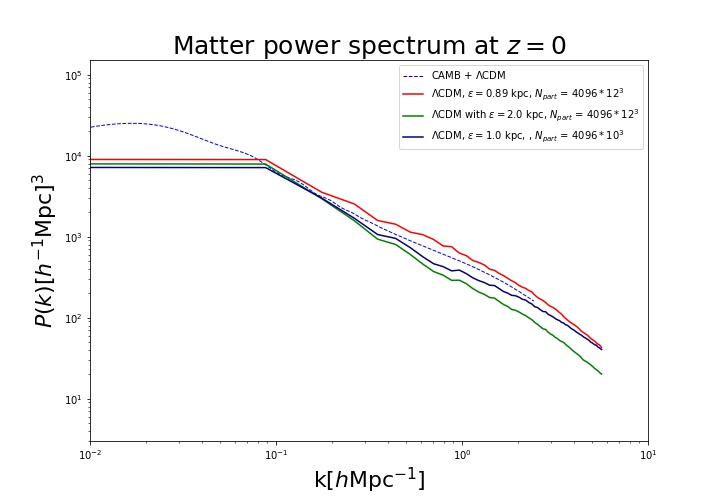}
    \caption{\footnotesize{Matter power spectrum as estimated by CAMB (dashed blue) and the 
    estimated by POWMES (solid lines) for this simulation for different softening lengths and number of particles.  
    The green line converges to a different value of $P(k)$ value meaning that the simulation is not creating the same amount of structure at these scales. The navy blue line has less particles and softening length of $\epsilon = 1$ kpc which reassembles the results to the red line on low scales.
    The constant part of the simulation is due the low resolution, which imposes a maximum scale 
    limit of the power spectrum (the lines are slightly displaced vertically for a better display).}}
    \label{fig7}
\end{figure}
To compute the mater power spectrum generated by the simulation, the code POWMES \cite{2009MNRAS.393..511C} 
comes in handy because it is designed to estimate the power spectrum of $N$-body simulations in 
an iterative form. The power spectrum $P(k)$ characterises the scales and clustering of galaxies 
in the Universe. In particular, many cosmological constrictions are based on the $P(k)$ 
measurement or its inverse Fourier transform, the two point correlation function.

\normalsize{
Figure (\ref{fig7}) shows the matter power spectrum computed by CAMB which is very 
close to the non-linear regime ($k \ll 1 $ indicates large scales) with a softening value 
of $\epsilon = 0.89$ kpc. The similarities are visible when comparing the solution 
of the Boltzmann equations and the numerical solution via the CDM simulation. 
As expected, the number of particles does affect the final result of the simulation; 
less particles lead to low matter power spectrum compared to the rest of the spectra,
as well as to the one computed by CAMB. 
It is also visible that when the softening parameter increases the simulation 
creates less structure, because is preventing the particles to come closer 
than 2 kpc. 
These are important parameters to bear in mind, as they need to be selected 
very carefully and effectively to get accurate results. 
The simple difference of one parameter can affect the whole result. 
}

\section{Conclusions and future work}

In this work we present a general description of the dark matter enigma, its discovery, and 
incorporation into the standard cosmological $\Lambda$CDM model. The model has been successfully 
tested through several observations, and compared with numerical simulations, in particular 
simulations at small scales where dark matter halos in galaxies are formed due to a 
spherical collapse model and on large scales by studying the cluster formation.

$N$-body simulations have been used in the cosmology field as an efficient tool to study process 
of large scale structure formation in the Universe. In this scheme, dark matter is 
modelled as a non-collisional fluid under the influence of a gravitational potential. 
Other models such as the 
SPH method involve gas dynamics for galaxy formation. Using these two methods, the interaction between dark matter particles and gas particles can be observed via the accretion of the gas into 
the dark matter halos. In this work, GADGET was used as a main code for simulating the large scale structure of a 50 Mpc Box in a $\Lambda$CDM Universe. The installation process of GADGET is also mentioned as an effective and simple guide to follow for young scientists.

The many parameters of an $N$-body code can be easily modified, and variations of them
can lead to a whole new physical system on the simulation. Thus it is important 
to test the various parameters and theories with a well established method (like the analytical 
solution to the Boltzmann equations) and then compare the variation of the parameters of the 
simulation. In this work, we performed different tests on various parameters which led to 
similar results predicted from an analytic $\Lambda$CDM model, but a variation on the number 
of particles as well as the softening length led to different results.
It is also important to highlight the key role that the gravitational softening plays on the 
simulations, different softening may recreate systems that resemble observations in different surveys. 

The data analysis requires the knowledge of statistical and probabilistic methods, and 
density distributions often studied in cosmology. Nevertheless, there exists some details 
within the structure formation theory that lead to issues in both observations and simulations, these issues are primarily the CUSP-CORE problem and the missing satellite problem. There are different alternatives to solve them, however, one of the main approaches in future work will be to compare different models to $\Lambda$CDM using different codes and initial conditions adapted for each one.

Several numerical codes were discussed, such as Tree, Particle-mesh, AMR and so on, listing their characteristics and differences. These codes are used for modelling diverse astrophysical systems, in particular, the $N$-body approach is used for galactic and cosmological systems,
mainly governed by the non-collisional Boltzmann equations.
The code N-GenIC comes very handy for generating initial conditions without gas particles, 
this software will be very useful to compare a GADGET modification which uses 
an axion model for dark matter particles called Axion-GADGET \cite{2018ApJ...853...51Z}. 
With this model, the main goal is to provide an alternative solution to several issues the 
$\Lambda$CDM model is dealing with, 
such as the CUSP-CORE problem \cite{1996ApJ...462..563N, 1997ApJ...490..493N, Moore:1994yx} 
observed on the density distribution of many galaxies and the missing satellite problem
\cite{1999MNRAS.310.1147M, 2005Natur.435..629S}. This modification continues in 
development and we are aiming to improve the  short range interaction between systems.

\section*{Acknowledgments}
J.A.V. acknowledges  FOSEC SEP-CONACYT Investigaci\'on B\'asica A1-S-21925 and 
UNAM-DGAPA-PAPIIT IA102219.

\bibliographystyle{ieeetr}
\bibliography{references}

\begin{thebibliography}{10}

\bibitem{2003itc..book.....R}
B.~{Ryden}, {\em {Introduction to cosmology}}.
\newblock 2003.

\bibitem{2007ApJS..170..377S}
D.~N. {Spergel}, R.~{Bean}, O.~{Dor{\'e}}, M.~R. {Nolta}, C.~L. {Bennett},
  J.~{Dunkley}, G.~{Hinshaw}, N.~{Jarosik}, E.~{Komatsu}, L.~{Page}, H.~V.
  {Peiris}, L.~{Verde}, M.~{Halpern}, R.~S. {Hill}, A.~{Kogut}, M.~{Limon},
  S.~S. {Meyer}, N.~{Odegard}, G.~S. {Tucker}, J.~L. {Weiland}, E.~{Wollack},
  and E.~L. {Wright}, ``{Three-Year Wilkinson Microwave Anisotropy Probe (WMAP)
  Observations: Implications for Cosmology},'' {\em \apjs}, vol.~170,
  pp.~377--408, June 2007.

\bibitem{1994ApJ...420..439M}
J.~C. {Mather}, E.~S. {Cheng}, D.~A. {Cottingham}, R.~E. {Eplee}, Jr., D.~J.
  {Fixsen}, T.~{Hewagama}, R.~B. {Isaacman}, K.~A. {Jensen}, S.~S. {Meyer},
  P.~D. {Noerdlinger}, S.~M. {Read}, L.~P. {Rosen}, R.~A. {Shafer}, E.~L.
  {Wright}, C.~L. {Bennett}, N.~W. {Boggess}, M.~G. {Hauser}, T.~{Kelsall},
  S.~H. {Moseley}, Jr., R.~F. {Silverberg}, G.~F. {Smoot}, R.~{Weiss}, and
  D.~T. {Wilkinson}, ``{Measurement of the cosmic microwave background spectrum
  by the COBE FIRAS instrument},'' {\em \apj}, vol.~420, pp.~439--444, Jan.
  1994.

\bibitem{1937ApJ....86..217Z}
F.~{Zwicky}, ``{On the Masses of Nebulae and of Clusters of Nebulae},'' {\em
  \apj}, vol.~86, p.~217, Oct. 1937.

\bibitem{1970ApJ...159..379R}
V.~C. {Rubin} and W.~K. {Ford}, Jr., ``{Rotation of the Andromeda Nebula from a
  Spectroscopic Survey of Emission Regions},'' {\em \apj}, vol.~159, p.~379,
  Feb. 1970.

\bibitem{1965ApJ...142..419P}
A.~A. {Penzias} and R.~W. {Wilson}, ``{A Measurement of Excess Antenna
  Temperature at 4080 Mc/s.},'' {\em \apj}, vol.~142, pp.~419--421, July 1965.

\bibitem{2017AJ....154...28B}
M.~R. {Blanton}, M.~A. {Bershady}, B.~{Abolfathi}, F.~D. {Albareti},
  C.~{Allende Prieto}, A.~{Almeida}, J.~{Alonso-Garc{\'{\i}}a}, F.~{Anders},
  S.~F. {Anderson}, B.~{Andrews}, and et~al., ``{Sloan Digital Sky Survey IV:
  Mapping the Milky Way, Nearby Galaxies, and the Distant Universe},'' {\em
  \aj}, vol.~154, p.~28, July 2017.

\bibitem{schutz:2009}
B.~Schutz, {\em {A First Course in General Relativity}}.
\newblock Cambridge University Press, 2nd~ed., June 2009.

\bibitem{1929PNAS...15..168H}
E.~{Hubble}, ``{A Relation between Distance and Radial Velocity among
  Extra-Galactic Nebulae},'' {\em Proceedings of the National Academy of
  Science}, vol.~15, pp.~168--173, Mar. 1929.

\bibitem{2013IJMPD..2230029M}
N.~{Mandolesi}, C.~{Burigana}, A.~{Gruppuso}, and P.~{Natoli}, ``{The Planck
  Mission: Recent Results, Cosmological and Fundamental Physics
  Perspectives},'' {\em International Journal of Modern Physics D}, vol.~22,
  p.~1330029, Dec. 2013.

\bibitem{1993ppc..book.....P}
P.~J.~E. {Peebles}, {\em {Principles of Physical Cosmology}}.
\newblock 1993.

\bibitem{10.1143/PTP.76.1036}
M.~Sasaki, ``{Large Scale Quantum Fluctuations in the Inflationary Universe},''
  {\em Progress of Theoretical Physics}, vol.~76, pp.~1036--1046, 11 1986.

\bibitem{1981PhRvD..23..347G}
A.~H. {Guth}, ``{Inflationary universe: A possible solution to the horizon and
  flatness problems},'' {\em \prd}, vol.~23, pp.~347--356, Jan. 1981.

\bibitem{2003RvMP...75..559P}
P.~J. {Peebles} and B.~{Ratra}, ``{The cosmological constant and dark
  energy},'' {\em Reviews of Modern Physics}, vol.~75, pp.~559--606, Apr. 2003.

\bibitem{1996ApJ...462..563N}
J.~F. {Navarro}, C.~S. {Frenk}, and S.~D.~M. {White}, ``{The Structure of Cold
  Dark Matter Halos},'' {\em \apj}, vol.~462, p.~563, May 1996.

\bibitem{1997ApJ...490..493N}
J.~F. {Navarro}, C.~S. {Frenk}, and S.~D.~M. {White}, ``{A Universal Density
  Profile from Hierarchical Clustering},'' {\em \apj}, vol.~490, pp.~493--508,
  Dec. 1997.

\bibitem{Moore:1994yx}
B.~Moore, ``{Evidence against dissipationless dark matter from observations of
  galaxy haloes},'' {\em Nature}, vol.~370, p.~629, 1994.

\bibitem{1999MNRAS.310.1147M}
B.~{Moore}, T.~{Quinn}, F.~{Governato}, J.~{Stadel}, and G.~{Lake}, ``{Cold
  collapse and the core catastrophe},'' {\em \mnras}, vol.~310, pp.~1147--1152,
  Dec. 1999.

\bibitem{2005Natur.435..629S}
V.~{Springel}, S.~D.~M. {White}, A.~{Jenkins}, C.~S. {Frenk}, N.~{Yoshida},
  L.~{Gao}, J.~{Navarro}, R.~{Thacker}, D.~{Croton}, J.~{Helly}, J.~A.
  {Peacock}, S.~{Cole}, P.~{Thomas}, H.~{Couchman}, A.~{Evrard}, J.~{Colberg},
  and F.~{Pearce}, ``{Simulations of the formation, evolution and clustering of
  galaxies and quasars},'' {\em \nat}, vol.~435, pp.~629--636, June 2005.

\bibitem{Matos:2008ag}
T.~Matos, A.~Vazquez-Gonzalez, and J.~Magana, ``{$\phi^2$ as Dark Matter},''
  {\em Mon. Not. Roy. Astron. Soc.}, vol.~393, pp.~1359--1369, 2009.

\bibitem{Magana:2012ph}
J.~Magana and T.~Matos, ``{A brief Review of the Scalar Field Dark Matter
  model},'' {\em J. Phys. Conf. Ser.}, vol.~378, p.~012012, 2012.

\bibitem{Hui:2016ltb}
L.~Hui, J.~P. Ostriker, S.~Tremaine, and E.~Witten, ``{Ultralight scalars as
  cosmological dark matter},'' {\em Phys. Rev. D}, vol.~95, no.~4, p.~043541,
  2017.

\bibitem{Mikkola1993}
S.~Mikkola and S.~J. Aarseth, ``An implementation ofn-body chain
  regularization,'' {\em Celestial Mechanics and Dynamical Astronomy}, vol.~57,
  pp.~439--459, Nov 1993.

\bibitem{1986Natur.324..446B}
J.~{Barnes} and P.~{Hut}, ``{A hierarchical O(N log N) force-calculation
  algorithm},'' {\em \nat}, vol.~324, pp.~446--449, Dec. 1986.

\bibitem{klypin1997particlemesh}
A.~Klypin and J.~Holtzman, ``Particle-mesh code for cosmological simulations,''
  1997.

\bibitem{2004astro.ph..3044O}
B.~W. {O'Shea}, G.~{Bryan}, J.~{Bordner}, M.~L. {Norman}, T.~{Abel},
  R.~{Harkness}, and A.~{Kritsuk}, ``{Introducing Enzo, an AMR Cosmology
  Application},'' {\em arXiv Astrophysics e-prints}, Mar. 2004.

\bibitem{springel2014high}
V.~Springel, ``High performance computing and numerical modelling,'' 2014.

\bibitem{chacon_2019}
J.~Chac\'on, ``Modelos de materia oscura: Una perspectiva num\'erica.'' Escuela
  Superior de F\'isica y Matem\'aticas, B.S. Thesis,
  \url{http://pelusa.fis.cinvestav.mx/tmatos/CV/3_RecursosH/Lic/Jazhiel_ESFM.pdf},
  Dec 2018.

\bibitem{parzen1962}
E.~Parzen, ``On estimation of a probability density function and mode,'' {\em
  Ann. Math. Statist.}, vol.~33, pp.~1065--1076, 09 1962.

\bibitem{1977MNRAS.181..375G}
R.~A. {Gingold} and J.~J. {Monaghan}, ``{Smoothed particle hydrodynamics -
  Theory and application to non-spherical stars},'' {\em \mnras}, vol.~181,
  pp.~375--389, Nov. 1977.

\bibitem{1997JCoPh.136..298M}
J.~J. {Monaghan}, ``{SPH and Riemann Solvers},'' {\em Journal of Computational
  Physics}, vol.~136, pp.~298--307, Sept. 1997.

\bibitem{1992ARA&A..30..543M}
J.~J. {Monaghan}, ``{Smoothed particle hydrodynamics},'' {\em \araa}, vol.~30,
  pp.~543--574, 1992.

\bibitem{2005MNRAS.364.1105S}
V.~{Springel}, ``{The cosmological simulation code GADGET-2},'' {\em \mnras},
  vol.~364, pp.~1105--1134, Dec. 2005.

\bibitem{1985A&A...149..135M}
J.~J. {Monaghan} and J.~C. {Lattanzio}, ``{A refined particle method for
  astrophysical problems},'' {\em \aap}, vol.~149, pp.~135--143, Aug. 1985.

\bibitem{2007nmai.conf.....B}
P.~{Bodenheimer}, G.~P. {Laughlin}, M.~{Rozyczka}, and H.~W. {Yorke}, eds.,
  {\em {Numerical Methods in Astrophysics: An Introduction}}, 2007.

\bibitem{2014MNRAS.444...62Y}
D.~{Yurin} and V.~{Springel}, ``{An iterative method for the construction of
  N-body galaxy models in collisionless equilibrium},'' {\em \mnras}, vol.~444,
  pp.~62--79, Oct. 2014.

\bibitem{2009MNRAS.394.1559G}
M.~{Grossi} and V.~{Springel}, ``{The impact of early dark energy on non-linear
  structure formation},'' {\em \mnras}, vol.~394, pp.~1559--1574, Apr. 2009.

\bibitem{2011MNRAS.415.2101H}
O.~{Hahn} and T.~{Abel}, ``{Multi-scale initial conditions for cosmological
  simulations},'' {\em \mnras}, vol.~415, pp.~2101--2121, Aug. 2011.

\bibitem{2006MNRAS.373..369C}
M.~{Crocce}, S.~{Pueblas}, and R.~{Scoccimarro}, ``{Transients from initial
  conditions in cosmological simulations},'' {\em \mnras}, vol.~373,
  pp.~369--381, Nov 2006.

\bibitem{1970A&A.....5...84Z}
Y.~B. {Zel'dovich}, ``{Gravitational instability: An approximate theory for
  large density perturbations.},'' {\em \aap}, vol.~5, pp.~84--89, Mar. 1970.

\bibitem{RevModPhys.61.185}
S.~F. Shandarin and Y.~B. Zeldovich, ``The large-scale structure of the
  universe: Turbulence, intermittency, structures in a self-gravitating
  medium,'' {\em Rev. Mod. Phys.}, vol.~61, pp.~185--220, Apr 1989.

\bibitem{1993MNRAS.262.1023W}
S.~D.~M. {White}, G.~{Efstathiou}, and C.~S. {Frenk}, ``{The amplitude of mass
  fluctuations in the universe},'' {\em \mnras}, vol.~262, pp.~1023--1028, June
  1993.

\bibitem{2000ApJ...538..473L}
A.~{Lewis}, A.~{Challinor}, and A.~{Lasenby}, ``{Efficient Computation of
  Cosmic Microwave Background Anisotropies in Closed Friedmann-Robertson-Walker
  Models},'' {\em \apj}, vol.~538, pp.~473--476, Aug 2000.

\bibitem{2009MNRAS.393..511C}
S.~{Colombi}, A.~{Jaffe}, D.~{Novikov}, and C.~{Pichon}, ``{Accurate estimators
  of power spectra in N-body simulations},'' {\em \mnras}, vol.~393,
  pp.~511--526, Feb. 2009.

\bibitem{2018ApJ...853...51Z}
J.~{Zhang}, Y.-L. {Sming Tsai}, J.-L. {Kuo}, K.~{Cheung}, and M.-C. {Chu},
  ``{Ultralight Axion Dark Matter and Its Impact on Dark Halo Structure in
  N-body Simulations},'' {\em \apj}, vol.~853, p.~51, Jan 2018.

\end{thebibliography}

\end{document}